\documentclass[a4paper]{article}
\usepackage[USenglish]{babel} %francais, polish, spanish, ...
\usepackage[T1]{fontenc}
\usepackage[ansinew]{inputenc}

\usepackage{lmodern} %Type1-font for non-english texts and characters

\usepackage{graphicx} %%For loading graphic files
\usepackage{amsmath}
\usepackage{amsthm}
\usepackage{amsfonts}

\begin{document}

\pagestyle{empty} %No headings for the first pages.

%% Title Page %%%%%%%%%%%%%%%%%%%%%%%%%%%%%%%%%%%%%%%%%%%%%%%
%% ==> Write your text here or include other files.

%% The simple version:
\title{Renormalization in  a Landau-to-Coulomb interpolating gauge in Yang-Mills theory}
%\author{A Andrasi}

\author{A Andra\v si$^+$ and J C Taylor\footnote{Corresponding author } 
\footnote{\textit{E-mail addresses} aandrasi@irb.hr (A. Andrasi), jct@damtp.cam.ac.uk (J. C. Taylor)} \\ \\ {\it  $^+$Vla\v ska 58, Zagreb, Croatia} \\ $^\dagger${\it DAMTP, Cambridge University, UK}}
%\date{} %%If commented, the current date is used.
%\affiliation{$^a$ Vla\v ska 58, Zagreb, Croatia\footnote[]{August 30, 2018}}
%\affiliation{$^b$ DAMTP, Cambridge University, UK}

%% The nice version:
%\input{titlepage} %%You need a file 'titlepage.tex' for this.
%% ==> TeXnicCenter supplies a possible titlepage file
%% ==> with its templates (File | New from Template...).

%% Inhaltsverzeichnis %%%%%%%%%%%%%%%%%%%%%%%%%%%%%%%%%%%%%%%
%\tableofcontents %Table of contents
%\cleardoublepage %The first chapter should start on an odd page.

%% Chapters %%%%%%%%%%%%%%%%%%%%%%%%%%%%%%%%%%%%%%%%%%%%%%%%%
%% ==> Write your text here or include other files.

%\input{intro} %You need a file 'intro.tex' for this.

\maketitle
%%%%%%%%%%%%%%%%%%%%%%%%%%%%%%%%%%%%%%%%%%%%%%%%%%%%%%%%%%%%%

%% ==> Some hints are following:

%\chapter{Some small hints}\label{hints}

\begin{abstract}

\noindent{The Coulomb gauge in QCD is the only explicitly unitary gauge. But it suffers from energy-divergnces which means that it is not rigorously well-defined. One way to define it unambiguously is as the limit of a gauge interpolating between the Landau gauge and the Coulomb gauge. This interpolating gauge is characterized by a parameter $\theta$ and the Coulomb gauge is obtained in the limit $\theta \rightarrow 0$. We study the renormalization of this $\theta$-gauge for all values of $\theta$, and note some special features of it.}\\

\noindent{Pacs numbers: 11.15.Bt; 03.70.+k}\\

\noindent{Keywords QCD, Coulomb gauge, renormalization}

\end{abstract}

\vfill\newpage

\pagestyle{plain} %Now display headings: headings / fancy / ...

\section{Introduction}
The Coulomb gauge in non-abelian gauge theory deserves attention  for  a number of reasons. It is the only explicitly unitary gauge. It may be convenient for the study of heavy bound states. It has been used in investigations of confinement, see for example \cite{cuc}, \cite{CH}, \cite{ZW} and references therein.

But, in perturbation theory at least, it suffers from 'energy divergences', that is Feynman integrals which are divergent  over the
time-components of the momenta, while the spacial components are held fixed.
The simplest such energy divergences occur at one loop. For pure YM theory, these are quite easily cancelled by combining Feynman
diagrams appropriately, but they are automatically removed by using the Hamiltonian, rather than the Lagrangian formalism \cite{MO}, so we use the Hamiltonian formalism in this paper.
%When quark loops are included, Ward identities secure the cancellation of this type of energy-divergence \cite{aajct1}.

More subtle, logarithmic, divergences appear first at two loop order. The cancellation of these was proved by Doust \cite{doust} (see also \cite{cheng}
and generalized in \cite{aajct1}).
The origin of these divergences was linked by Christ and Lee \cite{CL} with the problem of correctly ordering the factors in the
Coulomb potential in the Hamiltonian (see also \cite{SC}). But in this paper we consider only ordinary momentum-space Feynman perturbation theory,
in the manner of \cite{doust}.

 Pure energy-divergences occur
at 2-loop order only, not at higher order (see \cite{doust}). But if ordinary UV divergences are combined with energy divergences, new problems occur at 3-loop order, from the insertion of  UV divergent  loops into
two-loop  graphs \cite{aajct3}.
A difficulty in studying these matters is to be sure that the divergent integrals we are manipulating are well-defined.
To overcome this problem, we make use of a 'flow gauge', which interpolates between a covariant gauge
and the Coulomb gauge. A flow gauge is characterized by a parameter $\theta$, $\theta=1$ is the covariant
gauge, and the  Coulomb gauge is defined by the limit $\theta \rightarrow 0$. For nonzero $\theta$, there are no energy divergences
in any Feynman integral. 
In a previous paper \cite{aajct2}, we studied the flow gauge interpolating between the Feynman gauge and the Coulomb gauge. Let us call this the FC-flow gauge. The purpose of the present paper is to do the same for
a flow gauge interpolating between the Landau gauge and the Coulomb gauge: the LC-flow gauge.
This gauge has been used by some authors, for example \cite{ZW}.

As in \cite{aajct2}, we use the Hamiltonian (phase space) formalism, as this eliminates linear divergences to one-loop order. The renormalization of the Lagrangian formalism is obtained as a special case (see section 6 below).
The renormalization for the LC-flow gauge is obtained by field scaling and mixing, just as  in 
the FC-flow gauge \cite{aajct2}. In the present paper we derive the values of the renormalization constants
in the LC-flow gauge, which are in general different from those in the FC-flow gauge for $\theta\neq 0$.

In section 2 we define the LC-flow gauge, and state the Feynman rules. In section 3 we state the general form of renormalization, by scaling and mixing of fields and of sources. In section 4 we give the values of the renormalization constants to one-loop order, and in section 5 we show an example of their derivation.
In section 6 we discus the transition from the first order to the second order formalism.
Section 7 is a summary.

\section{The flow gauge}
The total Lagrangian density for the Hamiltonian form of Yang-Mills theory is
\begin{equation}
L_{TOTAL}=L+L_{GF}+L_G=L_C+L_{GF}+L_G+L_S,
\end{equation} 
%1
being the classical, gauge-fixing, ghost and source parts.
\begin{equation}
L_C=-E^{ai}F^a_{0i}-H
\end{equation}
%2
where $a$ is a colour index, $i$ is a 3-vector index, and
\begin{equation}
F^a_{0i}=\partial_0A^a_i-\partial_iA^a_0+gf^{abc}A^b_0A^c_i,
\end{equation}
%3
and $E^{ai}(x)$ (the 'colour electric field') is the conjugate momentum field to $A^a_i(x)$.

The Hamiltonian density is
\begin{equation}
H=-\frac{1}{2}E_i.E^i+\frac{1}{4}F_{ij}.F^{ij}
\end{equation}
%4
where we have suppressed colour indices, and use upper and lower indices although it is not covariant
(which explains the minus sign). The Lagrangian formalism would be obtained by expressing $L$ in terms of the free field
\begin{equation}
\tilde{E}_i=E_i-F_{0i}.
\end{equation}
%5

The gauge-fixing term which we use in this paper is
\begin{equation}
L_{GF}= -\lim_{\alpha \rightarrow 0}\frac{1}{2\alpha^2}(\partial_iA^i+\theta^2\partial_0A^0)^2.
\end{equation}
%6
The limit $\alpha \rightarrow 0$ imposes the constraint that the following bracket is zero.
The parameter $\theta$ dictates the flow, with $\theta=1$ giving the Landau gauge and $\theta=0$
the Coulomb gauge.

The terms involving the ghost $c$ and anti-ghost $c^*$ are
\begin{equation}
L_G=\partial_ic^*.[\partial^ic+gA^i\wedge c)]+\theta^2\partial_0c^*(\partial_0c+gA_0 \wedge c).
\end{equation}
%7
The source part of the Lagrangian is
\begin{equation}
L_S=
u^i.[\partial_ic+g(A_i\wedge c)]+u_0.[\partial_0c+g(A_0\wedge c)]+gv^i.(E_i\wedge c)
-\frac{1}{2}gK.(c\wedge c),
\end{equation}
%8
Here  $u^i, u_0,v^i,K$ are source fields used to impose the BRST
conditions. Except for $K$, these are fermionic (anti-commuting) like $c,c^*$. Colour indices are suppressed, and we use the abbreviation
\begin{equation}
(X\wedge Y)^a\equiv f^{abc}X^bY^c.
\end{equation}
%9
The BRST condition is
\begin{equation}
\int d^4x\left[\frac{\delta\gamma}{\delta u^i(x)}.\frac{\delta\gamma}{\delta A_i(x)}+\frac{\delta \gamma}{\delta u^0(x)}.\frac{\delta\gamma}{\delta A_0(x)}+
\frac{\delta\gamma}{\delta v^i(x)}.\frac{\delta\gamma}{\delta E_i(x)}+\frac{\delta\gamma}{\delta c(x)}.\frac{\delta\gamma}{\delta K(x)}\right]$$
$$ \equiv \gamma * \gamma =0,
\end{equation}
%10
where
\begin{equation}
\gamma = \int d^dy L(y).
\end{equation}
%11
 We will use  the notation that the momentum $k=(k_0,\textbf{K})$, K=|\textbf{K}|, $k^2=k_0^2-K^2$, and
\def\K{\bar{K}}
\begin{equation}
\K^2\equiv \textbf{K}^2-\theta^2 k_0^2.
\end{equation}
%12
\def\c{\gamma}
\def\t{\theta}
\def\R{\bar{R}}
\def\P{\bar{P}}
\def\Q{\bar{Q}}
We use $g_{\mu\nu}$ for the Minkowski metric with $g_{00}=1$. In the gauge given by (6), the Coulomb  propagator is
\begin{equation}
D_{00}=-\frac{K^2}{(\K^2)^2},
\end{equation}
%13
and the ghost  propagator is
\begin{equation}
 D= +\frac{1}{\K^2}.
\end{equation}
	
%14
The transverse part of the $A_i$
the  propagator is
\begin{equation}
D^T_{ij}=\frac{1}{k^2+i\eta}\left[g_{ij}+\frac{K_i K_j}{K^2}\right]
\end{equation}
%15
and the longitudinal part is
\begin{equation}
D^L_{ij}=-\frac{\t^4k_0^2}{(\K^2)^2}\frac{K_iK_j}{K^2},
\end{equation}
%16

Since we use the Hamiltonian formalism, we require also propagators involving the electric field $\bf{E}$. It is
\begin{equation}
D^{mn}=\frac{K^2}{k^2+i\eta}\left[ g^{mn} +\frac{K^m K^n}{\textbf{K}^2} \right],
\end{equation}
%17
which is transverse.
 (We use indices $i,j,...$ for the potential $\bf{A}$ and indices $m,n,...$ for $\bf{E}$).
There are also off-diagonal propagators. That between $E^m$ and $A_j$ has a transverse part
\begin{equation}
D_j^{Tm}=\frac{ik_0}{k^2+i\eta}\left[ g^m_j+\frac{K^mK_j}{\K^2} \right],
\end{equation}
%18
and a longitudinal part
\begin{equation}
D^{Lm}_j=i\t^2\frac{k_0}{\K^2}\frac{K^mK_j}{K^2}.
\end{equation}
%19
The off-diagonal propagator between $E_m$ and $A_0$ is
\begin{equation}
D^m_0=\frac{iK^m}{\K^2}.
\end{equation}
%20
Finally there is an off-diagonal propagator between $A_i$ and $A_0$
\begin{equation}
D_{0i}=-\frac{\t^2k_0K_i}{(\K^2)^2}.
\end{equation}
%21
In addition to the usual Yang-Mills vertex involving the spatial components $A_i$, there is an 
$E^{ma}A_0^bA_i^c$ vertex
\begin{equation}
-gf^{abc}g^i_m.
\end{equation}
%22
Most of the above rules are the same as for the FC-flow gauge, as in our previous paper \cite{aajct2}.
The three exceptions are $D_{ij}^L$ in (16), $D_j^{Lm}$ in (19) and $D_{0i}$ in (21).
(In general, these rules should be supplemented by a factor $[(2\pi)^4i]^{-1}$ for each propagator and a factor $(2\pi)^4i$ for each vertex; but these factors cancel  in 1-loop graphs.) 

Because of the constraint imposed by the gauge-fixing term (6), the propagators obey the following equations:
\begin{equation}
P^iD_{ij}+\t^2p^0D_{0j}=0,\,\,\,\,\,\,\, P^jD_{0j}+\t^2p_0D_{00}=0,
\end{equation}
%23
\begin{equation}
P^iD^n_i+\t^2p^0D^n_0=0.
\end{equation}
%24
 \section{Renormalization}
The structure of renormalization is as follows. Let the part $L$ of the Lagrangian, defined in (1), be the function
\begin{equation}
L(A_i,A_0,E^n;c,u^i,u^0,v_n,K;g).
\end{equation}
%25
This is the physical, finite Lagrangian. Then the renormalized (bare) Lagrangian is the function
\begin{equation}
L_R=
L(A_R^i,A_R^0,E_R^n,c_R,u_R^i,u_R^0,v_R^n,K_R;g_R),
\end{equation}
%26
where the renormalized fields and sources have the forms
\begin{equation}
A_R^i={Z'}_{5}^{1/2}A^i,\,\,\,\,A_R^0={Z'}_{6}^{1/2}A_0,\,\,\,\,c_R={Z'}_7^{-1/2}c,\,\,\,\,v^i_R={Z'}_{8}^{-1/2}v^i,$$
$$u^i_R={Z'}_5^{-1/2}\left[u^i+{Y'}_{10}\partial_0v^i-{Y'}_{11}v^i\wedge A^0\right],$$
$$u_R^0={Z'}_{6}^{-1/2}\left[u_0+{Y'}_{9}\partial_iv^i+{Y'}_{11}v^i\wedge A_i\right],$$
$$E^i_R={Z'}_{8}^{1/2}\left[E^i+{Y'}_{9}\partial^i A_0+Y'_{10}\partial_0 A^i+Y'_{11}A_0\wedge A^i+Y'_{12}v^i\wedge c\right],$$
$$K_R={Z'}_{7}^{1/2}\left[K+\frac{1}{2}Y'_{12}v^i\wedge v_i\right],\,\,\,\,g_R={Z'}_0^{1/2}g.
\end{equation}
%27
This structure was derived in our previous paper \cite{aajct2}, and applies to the Hamiltonian formalism of
gauge theories, irrespective of the gauge-fixing. But the values of the coefficients $Z'$ and $Y'$
depend on the gauge fixing, and are denoted by primed letters in the LC-flow gauge to distinguish
them from the FC-flow gauge. 
It can be proved (see \cite{FT} Appendix D, \cite{aajct2} Appendix B) that $L_R$ obeys the same BRST condition (10) as $L$, that is
\begin{equation}
\gamma_R * \gamma_R=0,
\end{equation}
%28
where the $*$ composition is defined in (10), and 
\begin{equation}
\gamma_R=\int d^dyL_R(y).
\end{equation}.
%29

The important point about (25) is that, for $E$,$u_i$ and $u_0$, there is mixing as well as scaling.
This implies that the renormalized Lagrangian in (26) does not have a pure Hamiltonian form:
it contains second order time derivatives of $A_i$.

 Given $L_R$ in (26), the renormalized ghost part, $L_{GR}$, is determined by the equation
\begin{equation}
\frac{\delta {\gamma_{GR}}}{\delta c^*(x)}=-{Z'_5}^{-1/2}\frac{\partial}{\partial x^i}\left[\frac{\delta \gamma_R}{\delta u_{iR}(x)}\right]-{Z'_6}^{-1/2}\t^2\frac{\partial}{\partial x_0}\left[\frac{\delta \gamma_R}{\delta u_{0R}(x)}\right],
\end{equation}
%30
where 
\begin{equation}
\gamma_{GR}=\int d^dy L_{GR}(y).
\end{equation}
%31
Then
\begin{equation}
L_{GR}=L_G(A_{iR}A_{0R},c_R,c^*_R,\theta_R;g_R),
\end{equation}
%32
where, in addition to  (27),
\begin{equation}
c^*_R={Z'}_5^{-1/2}c^*,\,\,\,\,\t_R^2=(Z'_5/Z'_6)^{1/2}\t^2.
\end{equation}
%33
Finally, the gauge-fixing term in  (6), $L_{GF}$, is unchanged by renormalization. If expressed in terms of the renormalized quantities (27), (33), it may be written
\begin{equation}
L_{GF}(A_i,A_0,\t,\alpha)=L_{GF}(A_{iR},A_{0R},\t_R,\alpha_R),
\end{equation}
%34
where
\begin{equation}
\alpha_R^2=Z_5\alpha^2.
\end{equation}
%35
Since $\alpha\rightarrow 0$, (35) makes no difference.

\section{Renormalization at one-loop order}
To one-loop order we write
\begin{equation}
{Z'}_5^{1/2}=1+a'_5, \,\,\,\,\,\,\, Y'_9=a'_9,
\end{equation}
%36
etc.   By calculating the divergent parts of a sufficient number of one-loop graphs, we have evaluated
the $a'$s. In terms of the quantity (using dimensional regularization with $d$ dimensions of space-time)
\begin{equation}
   c =\frac{g^2}{8\pi^2}C_G\frac{1}{4-d},
\end{equation}
%37
we find:
\begin{equation}
a'_0=a'_7=-\frac{11}{6}c,\,\,\,\,\,\,\,\,\, a'_{12}=0,
\end{equation}
%38
\begin{equation}
a'_5=\left[\frac{1}{2}-\frac{1}{12}\t+\frac{4}{3}\frac{\t}{1+\t}\right]c,\,\,\,\,\,\,\,\,\,
a'_6=\left[\frac{11}{6}-\frac{3}{4}\t\right]c,
\end{equation}
%39
\begin{equation}
a'_8=\left[-\frac{2}{3}+\frac{2}{3}\frac{\t}{1+\t}+\frac{1}{3}\frac{\t}{(1+\t)^2}\right]c,
\end{equation}
%40
\begin{equation}
a'_9=\left[\frac{1}{6}+\frac{2}{3}\t-\frac{2\t}{1+\t}+\frac{1}{3}\frac{\t}{(1+\t)^2}\right]c,
\end{equation}
%41
\begin{equation}
a'_{10}=\left[-\frac{1}{6}-\frac{1}{3}\t+\frac{2\t}{1+\t}-\frac{5}{3}\frac{\t}{(1+\t)^2}\right]c,
\end{equation}
%42
\begin{equation}
a'_{11}=\left[-\frac{1}{6}-\frac{2}{3}\t+\frac{2\t}{1+\t}-\frac{4}{3}\frac{\t}{(1+\t)^2}\right]c.
\end{equation}
%43

Some comments about these values may be useful. As expected, $a'_0$ is gauge-independent. The zero value of $a'_{12}$ and the equality between
$a'_0$ and $a'_7$ are consequences of equations (23) and (24) for the $v_j A_ic$ and $Kcc$ amplitudes respectively. 
Some combinations of constants have the same values as in the FC-flow gauge, for example
$a'_9+a'_{10}=a_9+a_{10}$. This is because the $v_ic$ graphs involve only the propagators (14), (18), (19) and (20) which are the same as in the FC-flow gauge. In  the Landau gauge ($\t=0$), $a'_5=a'_6=(13/12)c$,
 the expected values (see for example \cite{IZ} equation (12-113)).
In the Coulomb gauge ($\t=0$), $a'_9=-a'_{10}=-a'_{11}=1/6$, as found in the FC-flow gauge \cite{aajct2}.
We do not know the reason for this simplification.

We will now argue that the relation $a'_0=a'_7$ in (34) generalizes to all orders, that is
\begin{equation}
Z'_0=Z'_7.
\end{equation}
%44
Consider the renormalization of the $-gK.(c\wedge c)/2$ term in (8). The relevant divergent graphs
have two external ghost fields and a $K$ source.These graphs are overall logarithmically divergent
by dimensional power counting. (We assume that any divergencies in subgraphs have been cancelled by counter-terms).Take any one of the external ghost lines. It must attach to
a ghost-antighost-gluon vertex, say 
\begin{equation}
igp^ic^*(p).(A_i(p')\wedge c(q)),\,\,\,\,\text{or}\,\,\,\,  i\t^2gp^0c^*(p).(A_0(p')\wedge c(q)),
\end{equation}
%45
where the momenta are indicated, $q$ being an external momentum. The $A$ fields in the above equations are the ends of gluon propagators as in (15), (16), (19), (20), (21). Because of the properties (23) and (24),
$p'$ effectively gives zero, and so the internal momentum $p$ may be replaced by the external one $q$.
Then by power counting the graph is overall convergent. Thus the counter-terms available to cancel
the divergence must themselves cancel as in (38).

An example of the derivation of the above results (38)...(43) is given in the next section.

\section{An example}
To illustrate our methods, we describe the calculation of the divergent parts of the $A_0A_0$ amplitude.
The graphs involved are shown in Fig.1.
The individual graphs are, by power counting, quadratically divergent. We expect two powers of the external momentum to come out, reducing it to a logarithmic divergence. After that, the internal momenta
may be put equal to get the divergent part.
\begin{figure}
\centering
	\includegraphics[width=0.90\textwidth]{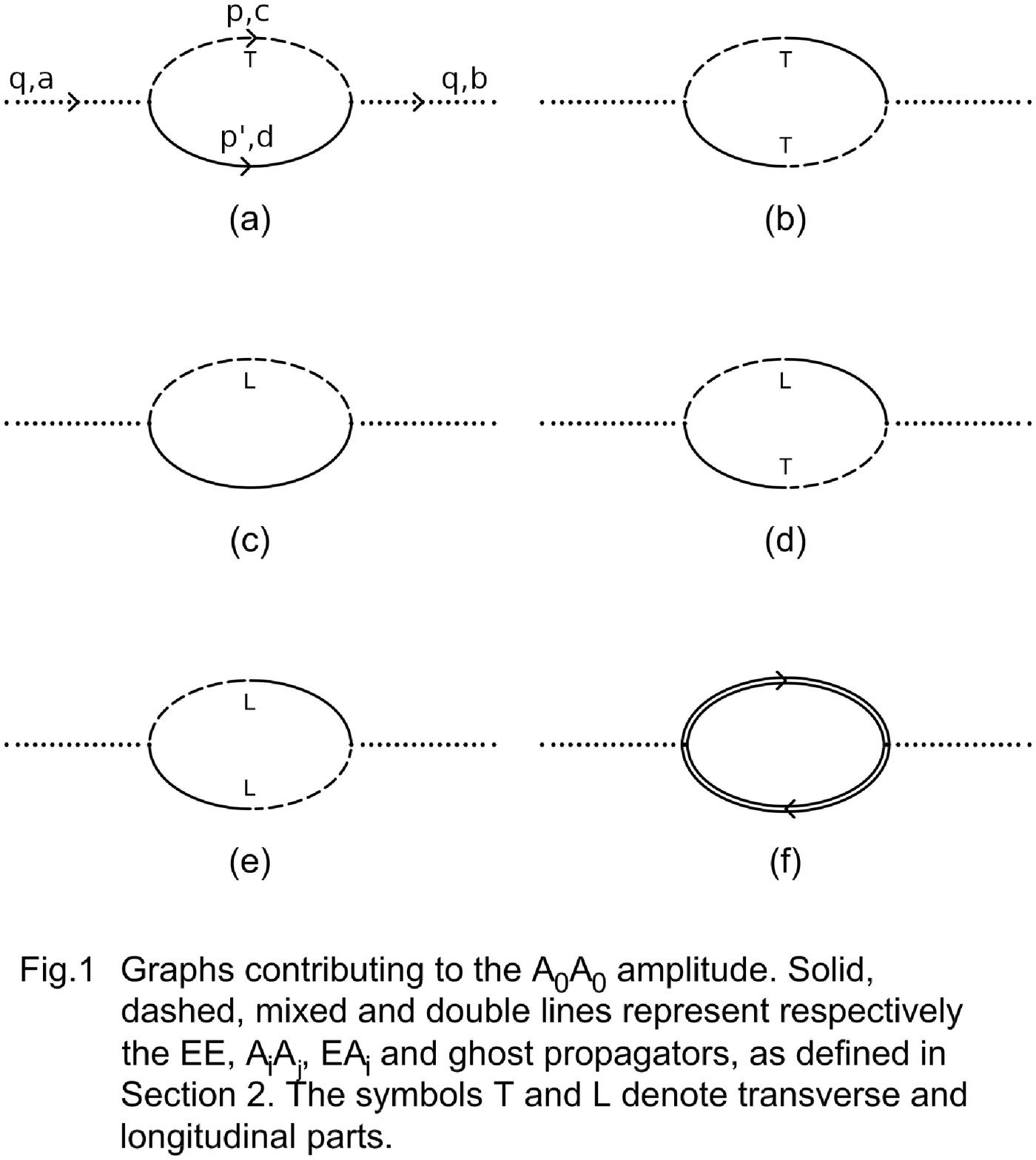}
	\label{fig:Fig1}
\end{figure}
Figures (a) and (b) are individually quadratically divergent, but, having carried 
out the $p_0$ integration, the sum gets a factor $(P-P')^2$, reducing the degree of divergence to logarithmic. The contribution is $icQ^2/3$.
This is the full result in the Coulomb gauge ($\t=0$).

In each of graphs (c) and (d), the transversality of the lower lines results in a factor $Q^2$.
For (c), this leads to the integral
\begin{equation}
g^2 C_G Q^2\frac{2}{3}\int d^{d-1}P\int dp_0 \frac{\t^4 p_0^2}{(\bar{P^2})^2p^2},
\end{equation}
%46
giving a divergent part
\begin{equation}
\frac{2}{3}\frac{\t^3}{(1+\t)^2}icQ^2.
\end{equation}
%47
Similarly (d) (together with the distinct graph $q\leftrightarrow -q$) give
\begin{equation}
-\frac{8}{3}\frac{\t}{1+\t}icQ^2.
\end{equation}
%48
There is partial cancellation between graphs (e) and (f), and the sum gives $(2/3)ic\t Q^2$
The sum of the above divergent parts determines the value of the counter-term $a'_9$ in (41).

\section{The transition to the Lagrangian formalism}
Before discussing the transition to the renormalized second order formalism, let us discuss the unrenormlized case, just using $L$ in (1). We can define  new variables
\begin{equation}
\tilde{E}_i=E_i-F_{0i}.\,\,\,\, \tilde{u}_0=u_0+D_iv^i, \,\,\,\,\, \tilde{u}_i=u_i-D_0v_i.
\end{equation}
%49
(Where $D_0=\partial_0+gA_0\wedge$ and $D_i=\partial_i+gA_i\wedge$ are covariant derivatives.)
In terms of this
\begin{equation}
L_C=\frac{1}{2}\tilde{E}_i\tilde{E}^i-\frac{1}{4} [F_{ij}F^{ij}+2F_{0i}F^{0i}],
\end{equation}
%50
%and it can be verified that, defining 
%\begin{equation}
%hat{L}_S=
%\tilde{u}^i.[\partial_ic+g(A_i\wedge c)]+\tilde{u}_0.[\partial_0c+g(A_0\wedge c)]+gv^i.(\tilde{E}_i\wedge c)
%-\frac{1}{2}gK.(c\wedge c),
%\end{equation}
%47
%\begin{equation}
%\int dx \hat{L}_S=\int dx L_S.
%\end{equation}
%48
%Thus the action in terms of the new variables (45) can be written jn the same functional form as
%in terms of the old ones (as in (8)).
The BRST condition (10) has the same form in terms of the new variables, that is
\begin{equation}
\gamma * \gamma=$$
$$\int d^4x\left[\frac{\delta\gamma}{\delta \tilde{u}^i(x)}.\frac{\delta\gamma}{\delta A_i(x)}+\frac{\delta \gamma}{\delta \tilde{u}^0(x)}.\frac{\delta\gamma}{\delta A_0(x)}+
\frac{\delta\gamma}{\delta v^i(x)}.\frac{\delta\gamma}{\delta \tilde{E}_i(x)}+\frac{\delta\gamma}{\delta c(x)}.\frac{\delta\gamma}{\delta K(x)}\right].
\end{equation}
%51
This is because the transformation to the new variables (49) is formally the same as a special case of  the transformation of the renormalized fields (27), which is known to satisfy (28).

Then, in (50), $\tilde{E}$ and $A$ are separated and the dependence on $\tilde{E}$ is trivial. However,
$\tilde{E}$ still occurs in $L_S$. But since $\tilde{E}$ occurs trivially in (50), we can omit the source $v_i$
in $L_S$. Then (51) still controls the BRST invariance of the second term in (50).

To make the transition to the second order formalism from the renormalized Lagrangian (26), we can proceed in the same way but using the renormalized fields in (27), finally obtaining
\begin{equation}
L_{CR}=
\frac{1}{2}\tilde{E}_{Ri}\tilde{E}^{i}_R-\frac{1}{4} [F_{Rij}F_R^{ij}+2F_{R0i}F_R^{0i}],
\end{equation}
%52
and
\begin{equation}
L_{SR}=u^i_R.[\partial_i c_R+g(A_{Ri}\wedge c_R)]+u_{R0}.[\partial_0 c_R+g(A_{R0}\wedge c_R)]
-\frac{1}{2}g_RK_R.(c_R\wedge c_R),
\end{equation}
%53
where the renormalized fields and sources are now just
\begin{equation}
A_R^i={Z'}_{5}^{1/2}A^i,\,\,\,\,A_R^0={Z'}_{6}^{1/2}A_0,\,\,\,\,c_R={Z'}_7^{-1/2}c,$$
$$u^i_R={Z'}_5^{-1/2}u^i,\,\,\,\,
u_R^0={Z'}_{6}^{-1/2}u_0,\,\,\,\,\,
K_R={Z'}_{7}^{1/2}K,\,\,\,\, g_R={Z'}^{1/2}_0g,
\end{equation}
%54
with the property (44) satisfied.
The renormalization constants $Z_8, Z_9, Z_{10}, Z_{11}$ are of course irrelevant to the second order formalism.

It may be noted that the source part of the action  has the same form in terms of the new variables (49)
as in terms of the original ones. That is, if we define
\begin{equation}
\tilde{L}_S=
\tilde{u}^i.[\partial_ic+g(A_i\wedge c)]+\tilde{u}_0.[\partial_0c+g(A_0\wedge c)]+gv^i.(\tilde{E}_i\wedge c)
-\frac{1}{2}gK.(c\wedge c),
\end{equation}
%55
then
\begin{equation}
\int d^dx \tilde{L}_S-\int d^dx L_S=$$
$$\int dx \left[ -v^i.(D_0D_i c)+v^i.(D_iD_0 c)+gv^i(F_{0i}\wedge c)\right]=0.
\end{equation}
%56
\section{Summary and comments}
We have studied, in Hamiltonian non-Abelian gauge theory, the flow gauge interpolating between the Landau and Coulomb gauges. We have described the general form of renormalization and calculated the one-loop renormalization constants as functions of the flow parameter $\t$. We make some comparisons with
the flow gauge interpolating between the Feynman and Coulomb gauges \cite{aajct2}.
We prove an all-orders equality between the ghost renormalization and the beta function (equation (44)),
which has previously been noted in \cite{ZW}. We demonstrate how the transition from the Hamiltonian to the Lagrangian formalism can be made.

The renormalization of the Hamiltonian form of gauge theories has a lot in common with that of
the covariant first-order formalism \cite{FT}, involving mixing as well as scaling of fields.
 
\section*{Acknowledgment}

We are grateful to the referee of our previous paper \cite{aajct2} for drawing our attention
to the Landau-to-Coulomb flow gauge.

\end{document}